\documentclass[a4paper,12pt]{article}
\usepackage[utf8]{inputenc}
\usepackage{cancel}
\usepackage{ulem}
\usepackage{amsfonts}
\usepackage{amssymb}
\usepackage{graphicx}
\usepackage{amsmath}
\usepackage{enumerate}
\usepackage{mathtools}
\usepackage{subfig}
\usepackage{color}
\usepackage{float}
\usepackage{here}
\usepackage{cite}
\usepackage{mathrsfs}
\usepackage{float,epsfig}
\usepackage{dcolumn}

\usepackage{graphicx}
\usepackage{bm}
\usepackage{amsmath,amssymb,amsthm}
\usepackage[colorlinks=true,linkcolor=blue,citecolor=red]{hyperref}
\newcommand{\be}{\begin{equation}}
\newcommand{\ee}{\end{equation}}
\newcommand{\bea}{\setlength\arraycolsep{2pt} \begin{eqnarray}}
\newcommand{\eea}{\end{eqnarray}}

\def\0{{\sst{(0)}}}
\def\1{{\sst{(1)}}}
\def\2{{\sst{(2)}}}
\def\3{{\sst{(3)}}}
\def\4{{\sst{(4)}}}
\def\5{{\sst{(5)}}}
\def\6{{\sst{(6)}}}
\def\7{{\sst{(7)}}}
\def\8{{\sst{(8)}}}
\def\sst#1{{\scriptscriptstyle #1}}

\title{
 \rightline{\mbox {\normalsize {
FISPAC-TH/19-31415,
UQBAR-TH/78-27182}}\bigskip}\bf \Large On Universal Constants   of  AdS Black Holes from
Hawking-Page Phase Transition }

\author{   Adil  Belhaj$^{1}$\footnote{belhajadil@fsr.ac.ma},  Anas El Balali$^{1}$,  Wijdane El Hadri$^{1}$,
 Emilio Torrente-Lujan$^{2,3}$\footnote{torrente@cern.ch}
	\hspace*{-8pt} \\
	{\small $^1$D\'{e}partement de Physique, \'Equipe des Sciences de la mati\`ere et du rayonnement,
		ESMAR}\\ {\small   Facult\'e des Sciences, Universit\'e Mohammed V de Rabat,  Rabat, Morocco} \\
	{\small $^{2}$Fisica T\'eorica, Dep. de F\'isica, Univ.  de Murcia,
Campus de Espinardo, E-30100 Murcia, Spain} \\  {\small  $^3$TH-division, CERN, CH-1211 Geneva 23, Switzerland}
}

\vspace{-3.6em}

\begin{document}
\maketitle
\begin{abstract}
We investigate the thermodynamic properties of the Hawking-Page phase transition of AdS black holes.
We  present evidence for the existence of  two   universal critical constants    associated with    the  Hawking-Page (HP) and minimum black hole thermodynamical transition points. These constants  are defined by    $C_S =\frac{S_{HP}-S_{min}}{S_{min}}$ and  $C_T =\frac{T_{HP}-T_{min}}{T_{min}}$ where  $S_{min}(S_{HP})$
 and  $T_{min}(T_{HP})$ are the minimal (HP phase transition)  entropy and temperature, respectively, below which no
black hole can exist.
 For a large class of   four dimensional non-rotating black holes, we find $C_S =2 $ and  $C_T = \frac{2-\sqrt{3}}{\sqrt{3}}$.
  For the rotating case, however, such  universal  ratios  are slightly affected without losing the expected values.
   Taking small values of the  involved rotating parameter, we  recover the same  constants.  Higher dimensional models, 
    with other universal constants,
     are  also discussed in some details. 

{\noindent}

\end{abstract}
\newpage


\section{Introduction}

A classical stationary black hole (BH) solution in General Relativity is characterised
by its mass $M$, angular momentum $J$ and charge $Q$ alone. In particular, its horizon area is a
simple function of these three quantities. Identifying the horizon area as an entropy,  the BHs  satisfy  a set of laws  which are
directly analogous to those of thermodynamics \cite{BCH,H1,H2}.
According to Hawking’s prediction \cite{H1,H2},  BHs
 emit thermal radiation at the semiclassical level which  fixes  the Bekenstein-Hawking
area/entropy relation to be $S_{BH} \sim \frac{A}{4}$  (where  $c = \hbar = G = 1$).
The thermodynamic interpretation of BH physics  might be more than a mere analogy and  could provide  the existence of sublying degrees of freedom\cite{SV}.
Any  consistent  theory of quantum gravity  could  address this challenge in
some way or at least advance in this sense.  More detailed studies of  black holes could be found in many papers including  \cite{I1,I3}.
 The thermodynamic properties of anti de Sitter
 (AdS) black holes have been extensively
studied, which  includes the existence of a  minimal Hawking temperature and
the Hawking-Page phase transition \cite{w2,w3}.
The Hawking-Page phase transition between large stable black holes and thermal
 gas in the AdS space  has been approached  using different methods. For example,  an analogy between phase structures of various AdS black holes
 and statistical models associated with Van der Waals like phase transitions
has been suggested\cite{I9}.
 Interpreting the cosmological constant   as a kind of  thermodynamic pressure,
and its conjugate  variable as the thermodynamic volume,   several
non trivial results have been presented  \cite{w4,I11}.

Thermodynamics of  AdS black holes, in supergravity theories,  have been
also investigated by exploiting  the  AdS/CFT correspondence  which provides an
 interplay  between gravitational models in $d$-dimensional AdS geometries
and $(d-1)$-dimensional conformal field theories living in the  boundary of
such AdS spaces.  Using  the physics  of solitonic  branes,  different models in
type IIB superstrings and M-theory have been studied by considering the
 cosmological constant in the bulk closely related to the number of colors
associated with branes in question. The thermodynamic stability behavior of such
AdS black holes, in higher dimensional known supergravity theories has
been examined in this context \cite{I18,X0}.


More concretely, at the semiclassical level, it has been shown that
 AdS black holes can be in  stable thermal equilibrium with radiation.
Hawking and Page  presented  a theoretical   evidence for the existence
of  certain phase transitions in the phase space of the (non-rotating uncharged)
Schwarzschild-AdS black hole \cite{1}. Later, a 
 first order phase transition in the charged (non-rotating)
Reissner–Nordstrom-AdS (RN-AdS) black hole space-time has been investigated \cite{3,4,5}.
Moreover,
  many efforts have been devoted to deal with  various AdS black holes in different backgrounds including  branes and  Dark Energy (DE)
\cite{6,7,8,9,10,11,12,13}.

A close inspection  in  thermodynamic  behaviors of AdS black holes reveals  that  there exist
two  temperatures. The first  one   is the
 minimum temperature $T_{min}$   below which no
black hole can survive. The second one is called  Hawking-Page (HP)
 temperature $T_{HP}$ such that
for $T<T_{HP}$ the thermal  AdS is the preferred state while for
$T>T_{HP}$  the black-hole is  the preferred one: the one with the dominant contribution to the
semiclassical partition function. This
 is the Hawking-Page  phase  transition \cite{10}.

According to the  usual,  albeit
rather qualitative, wisdom, the Hawking-Page transition indicates
that theories must have a small number of states at lower energy, but a huge  number of states
at high energies with a sharp transition at $T_{HP}$.
 It turns out that such  temperatures could be   combined to  reveal certain universal behaviors.
 Such  an issue  has  been
   unveiled in a previous work  \cite{14}  and  similar ideas have  been discussed in \cite{15}. Moreover,
   universalities
   in AdS black holes have been approached  from other angles \cite{9}.  This could open a
    new window for investigating
   universal  behaviors   which could   be considered as tools for  understanding  physical models
   of black holes.

 The aim of this work is   to  quantify in a precise way
the jump in the number of states at the
 Hawking-Page  transition by  showing  how some kind of universality behavior is at work.  Precisely, we   prove
    two   universal critical constants    associated with  thermodynamic  transitions of four dimensional  AdS black holes.
     They are given by    $C_S =\frac{S_{HP}-S_{min}}{S_{min}}$ and  $C_T =\frac{T_{HP}-T_{min}}{T_{min}}$  where  $S_{min}$  and  $S_{HP}$
 are  the minimal    and    $HP$  phase  entropies, respectively.
 For a large class of non rotating black holes, we find $C_S =2 $ and  $C_T = \frac{2-\sqrt{3}}{\sqrt{3}}$.
 For the rotating solution,  however, such  universal ratios are slightly affected without losing the expected values.
  For small values of the  involved rotating parameter, we  recover the same  constants. Higher dimensional solutions, with other universal constants,  are also examined in some details.  In this work, we use
dimensionless units in which one has  $\hbar= G_4= c = 1$.

 This paper  is  structured as  follows. In section 2, we give the strategy of this work.
  In section 3,  we deal with the non-rotating  AdS black hole  by computing the associated critical ratios.
   In section  4,  we investigate the  rotating AdS black holes.  Concluding  discussions and open questions  are presented in section 5.

\section{ AdS black hole universalities  in the $(S,T)$ plane}

Many   constants arise in the formulation of fundamental physical  theories.  Such
 invariant quantities, called fundamental constants, are considered as scalar ones in any  coordinate system.
    The dimensionless type of such  universal constants
   has been  of a particular interest. In this way, these constants are expressed as simple numbers with non-trivial meanings.
   Thus, theirs   determinations  are very important. With the help of experimental
evaluations, one  could  predict wether a theory is relevant or not. Besides, the universal constant
 values have been   essential for a precise quantitative description of fundamental theories of the universe. In this paper, we
   investigate   critical  behaviors
of  AdS black holes  by proposing  some  universal ratios and constants  associated  with  the Hawking-Page phase transition
  from  laws of thermodynamics by varying certain parameters\cite{16}. The  variation of
    such  parameters  have been considered  from many reasons. For instance,  this could provide   a  possible origin of such parameters from
     ‘more fundamental’ theories. Moreover, they  could be supported by
 the need of keeping the scale law (the first degree homogeneity) of the
"internal energy" with respect the extensive thermodynamic variables
(the consistency of the first law of black hole
thermodynamics with the Smarr relation) \cite{11}.
Before going ahead,  we remind  some black hole thermodynamics.
In the simplest case, the Gibbs (or  rather the Helmholtz) function is defined by
\bea
G&=& M- T S,
\eea
where $M$ is the total "energy" or mass of the system, which obeys the "first law
 of black hole thermodynamics" and
where the temperature is given by
\bea
T &=& \frac{\partial M}{\partial S}.
\eea
More  generally, the Gibbs function is a function of an intensive variable
the temperature and any other extensive magnitudes
\bea
 G=G(T,X_i)
 \eea
where $X_i$ denote such  extensive magnitude needed  to describe the system.

\begin{figure}
\begin{center}
\includegraphics[scale=0.5]{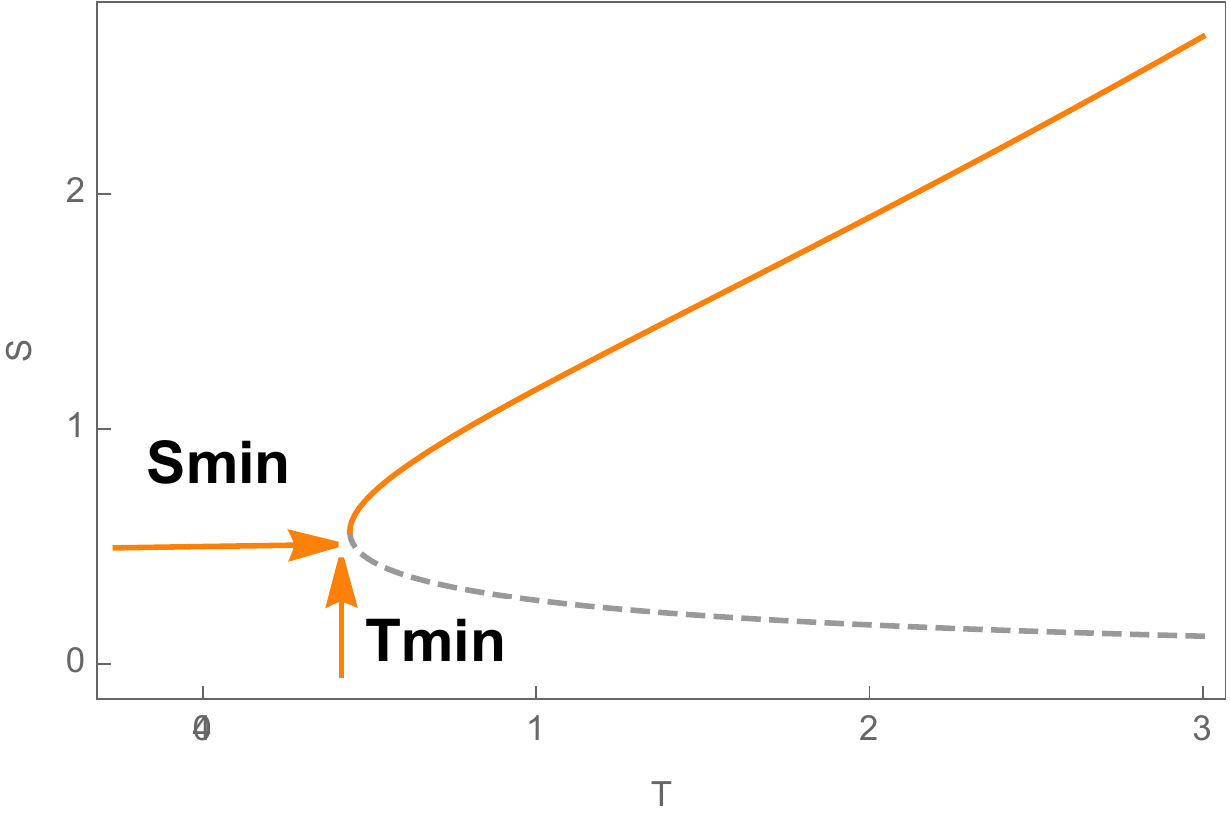}\hspace{1.2cm}\includegraphics[scale=0.5]{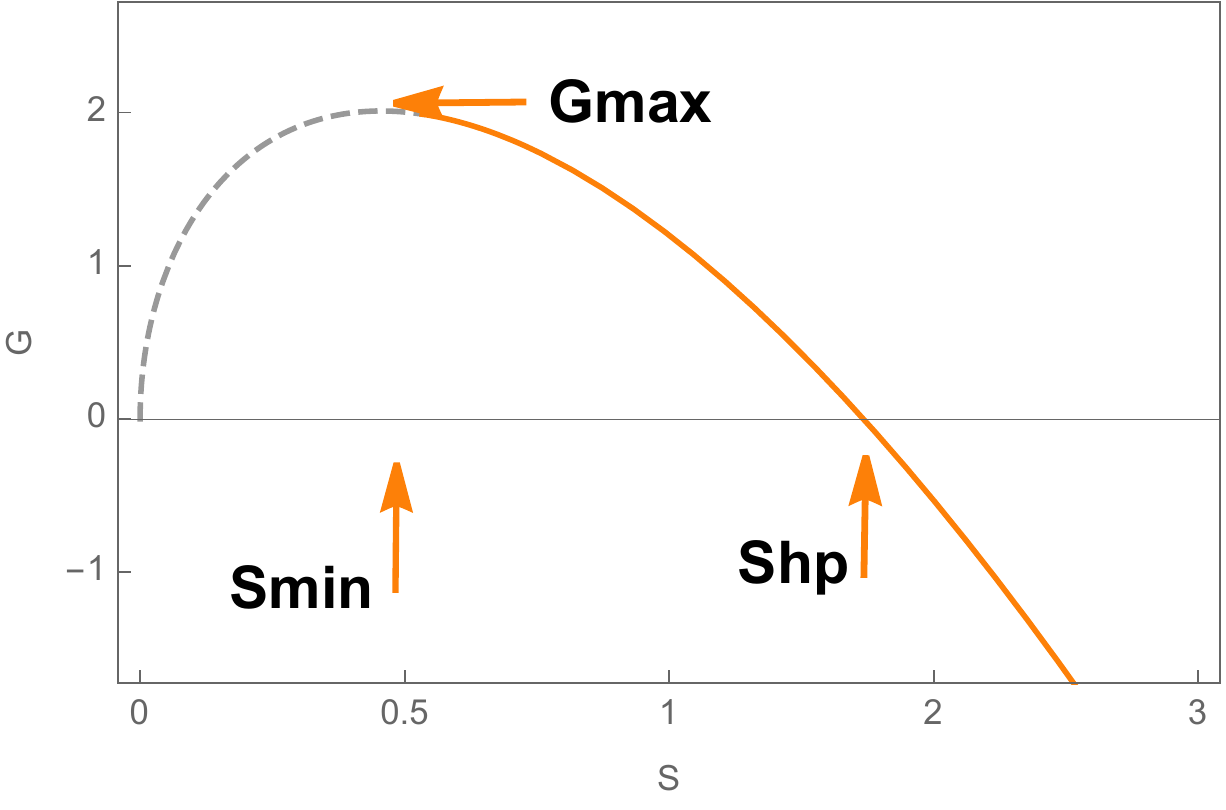}
\caption{ Qualitative behavior (arbitrary units).: (Left) S-T plane, (Right) G-S plane.}
\label{fig2}
\end{center}
\end{figure}

In standard macroscopical thermodynamics, the third law  says  that
the entropy of a system approaches a constant value, $S_{min}$ as its temperature approaches absolute zero. This constant
 value is absolute which cannot depend on
any other parameters characterising the closed system. Moreover,  at absolute zero the system must be in a state of minimum energy.
In the simplest case,  there is only an unique state ($S_{min}\equiv 0$).
In other systems, however,  there may remain some finite entropy, either because the system becomes locked in a non-minimal
 energy state, or, because the minimum energy state is not unique.
This constant value, $S_0$ is usually called the "Residual Entropy" of the system, which  occurs if a system can exist
in many different states at lowest energies
\bea
\lim_{T\to T_{min}} S(T)&=& S_{min},\label{e001}\\
\lim_{T\to T_{min}} \left (\frac{ \partial S}{\partial X}\right)_T &=& 0.
\eea
The condition \eqref{e001} is equivalent to the  constraint
\bea
\left .\frac{\partial T}{\partial S} \right |_{S_{min}}=
 \frac{\partial^2 M}{\partial S^2} =0.
\eea
Moreover, it is recalled  that
\bea
\frac{\partial G}{\partial S}&=& \frac{\partial U}{\partial S} -\frac{\partial T}{\partial S} S-T
 =-\frac{\partial T}{\partial S} S.
\eea
If we take $S=S_{min}$ ( $\partial T/\partial S=0$), then in this case one has  $\frac{\partial G}{\partial S}=0$.
 As illustrated in Fig.\ref{fig2}, the  free energy reaches a maximum  at this point
\bea
G(S_{min}) &=& G_{max}.
\eea
In the present work,  we will compute the minimal  or the  residual Entropy for a large class of AdS  black hole systems
 and  relate it to the Hawking-Page entropy $S_{HP}=S(T_{HP})$ constrained by
\bea
G(T_{HP},X_{fix}) &=&0.
\eea
It should be noted that  the entropy in general and the particular values  $S_{HP}$ and $S_{min}$ depend on any
other parameters. It has been  understood that both are computed at the same value of them.
We demonstrate here some universal relations and constants associated with the $HP$ phase transition.

As it was indicated in the introduction, the Hawking-Page transition indicates that quantum theories of gravity
 must have a small number of states at lower energy, but a huge  number of states at higher energies with a sharp transition at $T_{HP}$.
   The aim of this work is   to  quantify in a precise way
the jump in the number of states at the
 Hawking-Page  transition by  showing  how some kind of universal behavior is at work.

  The main aim, through this work,   is to investigate
  two universal constants at the Hawking-Page transition point. Precisely, they are given by
\bea
C_S &=&\frac{S_{HP}-S_{min}}{S_{min}}\\ C_T &=&\frac{T_{HP}-T_{min}}{T_{min}}.
\eea
In particular, we will show that the numerical values,   for non-rotating black holes,  are
\bea
C_S &=& 2 \\ C_T &=& \frac{2-\sqrt{3}}{\sqrt{3}}.
\label{twocostantes}
\eea
Equivalently, they   can be given also  by
\bea
S_{HP} &=& 3  S_{min},\\
T_{HP} &=& \frac{2\surd 3}{3} T_{min}.
\eea
being  independent of any parameter. For rotating  AdS black holes, however,     these ratios depend on other thermodynamic 
  parameters. They can be put like
\bea
S_{HP} &=& 3  S_{min}+....,\\
T_{HP} &=& \frac{2\surd 3}{3} T_{min}+....
\eea
 In what follows, we illustrate such results by giving explicit models with AdS backgrounds. We expect
 that these explored universal constants  could be exploited  to unveil more  physical data associated with thermodynamic
 aspects of  AdS black holes in four dimensions.

 Let us remark that  the systematic study of other possible universal relations involving other parameters than
  $T$ and $S$  is beyond the scope of this work.

   However, let us simply remark here that
  we could  calculate other constants associated with other parameters including  angular momentum $J$ or the charge $Q$ of the black hole.  In the extended phase space $(P,V)$, for instance, it has been  observed  that  there exists  an  universal constant involving   the volume which  will be   given   in the conclusion part. However, no universal constants corresponding to   the pressure can be elaborated  because   such  phase transitions can occur at all pressure values.  This also holds for transitions of  the charged and rotating black holes which  can happen at generic  values of  the electric potential and the angular velocity.

\section{ Universal constants of non-rotating AdS-black holes}

In this section, we investigate  non rotating  AdS black holes in  four dimensions. Concretely, we  first consider a
 Schwarzschild solution. Then, we deal with  a  charged  black hole.
\subsection{Schwarzschild AdS black holes}
We first consider the simplest case of a  non charged AdS-black hole  defined by the following action
\begin{equation}
I=\frac{1}{16 \pi} \int_{\mathcal{M}}d^4 \, x \sqrt{-g} \left( R -2 \, \Lambda \right)
\end{equation}
where $\Lambda$ is the cosmological constant. According   to  \cite{10}, this theory involves a black hole solution given by the following  line element
\begin{equation}
ds^2=-f(r)dt^2+f^{-1}(r)dr^2+r^2 \,d\Omega^2,
\label{metric1}
\end{equation}
where $d\Omega^2$ is  a 2-dimensional  unit sphere.  $
f(r) =1-\frac{m}{r} + \frac{r^{2}}{L^{2}}$  denotes the metric function
where  $L$ is  a fixed constant, considered  as the  AdS radius $L$  being related  to  a  negative   cosmological constant,
via  $\Lambda=-3/L^2$. Here, we consider two cases associated with such a  constant.\\
{\bf Case 1: Fixed cosmological constant}\\
Using a normalised  entropy $S\equiv \frac{ S_{BH}}{\pi}=\frac{A}{4\pi}= \, r_h^2$ with $r_h$ the radius of the black hole
 horizon and taking $L=1$, the black hole mass is
\begin{equation}
M=\frac{\sqrt{S} \left( 1 + S \right)}{2}.
\label{msch}
\end{equation}
The temperature  has the following form
\begin{equation}
T=\frac{\partial \,M}{\partial \, S}=\frac{1 +3S}{4  \sqrt{S}}.
\label{tsch}
\end{equation}
The variation of the temperature gives   the minimal entropy and the minimal temperature $\left( S_{min},T_{min} \right)$. A simple calculation shows that
\begin{equation}
\left( S_{min},T_{min} \right)=\left(\frac{1}{3}, \frac{\sqrt{3}}{2} \right).
\label{Sschmin}
\end{equation}
For  the  Hawking-Page phase transition, one should use  the  Gibbs free energy  given by the following formula
\begin{equation}
G= M -T \cdot S=\frac{1}{4} \sqrt{S} (1-S).
\label{Gsch}
\end{equation}
It is easy to find
\begin{equation}
\left( S_{HP},T_{HP} \right)=\left(1,1\right).
\label{Sschhp}
\end{equation}
Combining Eq \eqref{Sschmin} and \eqref{Sschhp}, we  obtain the desired  universal critical ratios
\begin{equation}
\frac{S_{HP}}{S_{min}}={3}, \quad \frac{T_{HP}}{T_{min}}=\frac{2}{\sqrt{3}}.
\label{without}
\end{equation}
{\bf Case 2:  Thermodynamical cosmological constant}\\
Treating  the cosmological constant as a thermodynamical variable playing the role of a pressure $
\Lambda=-\frac{3}{L^2}=-8 \pi P$,
the  first law of BH thermodynamics becomes
\begin{equation}
dM=TdS+VdP,
\end{equation}
where $V$ is the thermodynamic volume conjugated to the pressure $P$ ($V=\partial M/\partial P$) \cite{17}.
In this case, the masse takes the following form
\begin{equation}
M=\frac{\sqrt{S} \left( 3 + 8 \pi S P \right)}{6}.
\label{msch}
\end{equation}
The temperature and  the volume are  given by
\bea
T&=&\frac{1 +8 P \pi S}{4  \sqrt{S}}\\
V&=&\frac{4}{3} \pi S^{\frac{3}{2}}.
\label{tsch2}
\eea
 An easy   computation  reveals  that the minimal temperature (at fixed $P$) and the associated entropy  are
\begin{equation}
\left( S_{min},T_{min} \right)=\left(\frac{1}{8 \pi P}, \sqrt{2 \pi P} \right).
\label{Sschmin2}
\end{equation}
For this model, the   canonical Gibbs free energy reads as
\begin{equation}
G= M -T \  S =\frac{\sqrt{S}}{12}  (3-8 \pi P S).
\label{Gsch2}
\end{equation}
Using the normalised entropy, we find  the critical point in  the $\left(S,T \right)$ space
\begin{equation}
\left( S_{HP},T_{HP} \right)=\left(\frac{3}{8\pi P}, \sqrt{\frac{8 \pi P}{3}} \right).
\label{Sschhp2}
\end{equation}
Combining Eq \eqref{Sschmin2} and \eqref{Sschhp2}, we  find $\frac{S_{HP}}{S_{min}}={3}$ and  $\frac{T_{HP}}{T_{min}}=\frac{2}{\sqrt{3}}.$
\subsection{Reissner–Nordstrom-AdS (RN-AdS) black holes}
 This four dimensional AdS black hole  is defined by the metric function 
\begin{equation}
f(r)=1-\frac{m}{r}+\frac{r^{2}}{L^{2}}+\frac{Q^{2}}{r^{2}},
\label{fRN}
\end{equation}
where $Q$ is the charge \cite{10}.
Using  a  entropy  reduced expression and  putting $L=1$,  the black hole mass  reads as
\begin{equation}
M=\frac{\sqrt{S} \left(S+  \left(1+\Phi^2 \right) \right)}{2},
\end{equation}
where $\Phi=\frac{Q}{r_+}$ is  the electric potential conjugated to the charge   which  can be defined as the difference between
 the electric potential at the event horizon and the boundary $r \rightarrow \infty$.   For simplicity reasons,
  we will consider only a thermodynamic  phase space   corresponding to 
the variables $(S,Q)$  by  fixing  $L$.  It is worth to note that   the  extension  to a  more general case, where the   latter
is also  variable,  could be possible as in the  non-charged case.
Using the associated  first law of thermodynamics,  the expression of the temperature can be written as
\begin{equation}
T=\frac{3S +  \left( 1 - \Phi^2 \right)}{4 \sqrt{S}}.
\end{equation}
The variation of the temperature with respect to the entropy gives
\begin{equation}
\left( S_{min}, T_{min} \right)= \left( \frac{ \left( 1 - \Phi^2 \right)}{3}, \frac{\sqrt{3}}{2} \left( 1 - \Phi^2 \right)^{1/2} \right).
\label{RNMIN}
\end{equation}
To get the remaining quantities corresponding to   the Hawking-Page phase transition,  one can exploit  the following grand   canonical Gibbs free energy
\begin{equation}
G=\frac{\sqrt{S} \left[ \left(1-\Phi^2 \right) - S\right]}{4 }.
\end{equation}
Imposing $G(T_{HP})=0$,  we find
\begin{equation}
\left( S_{HP}, T_{HP} \right)= \left( \left( 1 - \Phi^2 \right),  \left( 1 - \Phi^2 \right)^{1/2} \right).
\label{RNHP}
\end{equation}
Using Eq.\eqref{RNMIN} and Eq.\eqref{RNHP},  we  can determine  the expressions of the universal critical ratios.
After a simple examination of such  the involved relations,  we find  $C_S=2$ and  $C_T=\frac{2-\sqrt{3}}{\sqrt{3}}$.
 It is  interesting to precise that  one can also
  consider other non-rotating $AdS$ black holes by implementing other external parameters including   non-trivial  backgrounds.
   We expect that such models
    provides similar universal critical constants.\\
Having discussed  explicit models for non-rotating AdS models, we move to  consider  a rotating model. We will show that
 the latter exhibits identical universal ratios by considering small values of the involved parameters.

\section{Rotating AdS black holes and universalities}
Kerr black holes are characterised by their mass  and angular momentum. The latter provides non trivial behaviors, even for optical aspects.
Following \cite{18,19}, the line element of a   four dimensional  Kerr-AdS metric  reads as
\begin{equation}
\begin{aligned}
ds^{2}= \frac{\Sigma^{2}}{\Delta_{r}}dr^{2}+\frac{\Sigma^{2}}{\Delta_{\theta}}d\theta^{2}+\frac{\Delta_{\theta}sin^{2}\theta}
{\Sigma^{2}}(a\frac{dt}{\Xi}  -(r^{2}-a^{2})\frac{d\phi}{\Xi})^{2}  -\frac{\Delta_{r}}{\Sigma^{2}}(\frac{dt}{\Xi}-a\sin^{2}\theta
\frac{d\phi}{\Xi})^{2}.
\end{aligned}
\end{equation}
The involved terms are
\begin{equation}
\begin{aligned}
& \Delta_{r}=r^{2}-2mr+a^{2}+\frac{r^{2}}{L^{2}}(r^{2}+a^{2}), ~~~~ \Delta_{\theta}=1-\frac{a^{2}}{L^{2}}\cos^{2}\theta, \\
& \Xi=1-\frac{a^{2}}{L^{2}}, \qquad \qquad \qquad \qquad \qquad  ~~~~ \Sigma^{2}=r^{2}+a^{2}\cos^{2}\theta.
\end{aligned}
\end{equation}
Using  the normalized entropy $(S\equiv \frac{ S_{BH}}{\pi})$ and the scaled  quantities
\begin{eqnarray}
\Omega&=&\frac{J}{MS}\left(1+S\right),  \label{OS}\\
\frac{J}{S}&=&\frac{M\Omega}{1+S}, \label{JS}\\
a&=&\frac{J}{M}=S\Omega\left(1+S\right)^{-1}.\label{12}
\end{eqnarray}
 where   $\Omega$  is  the difference between the angular velocities at the event horizon ($\Omega_h$) and at infinity
  ($\Omega_{\infty})$, one obtains    the  mass  which is  given by
\begin{equation}
M^2 =\frac{S}{4}\frac{\left(1+S \right)^2}{1-\frac{\Omega^2 S}{1+S)}}.
\label{Msquare}
\end{equation}
Similarly,  the Hawking temperature  is found to be
\begin{equation}
T=\sqrt{\frac{S(1+S)}{(1+S-S\Omega^2)}}\left[\frac{1-2 S(\Omega^2-2)-3S^2(\Omega^2-1)}{4S(1+S)}\right].
\label{TKERR}
\end{equation}
Varying the temperature   with respect to the  normalised entropy,   we find  the  minimal  entropy
\begin{equation}
S_{min}= -\frac{2}{3} \left( \frac{ 1 - \Omega^2/2 }{  1 - \Omega^2  } \right)
 + \left( \lambda_1+ \lambda_2 \right),
 \label{Skerrmin}
\end{equation}
where the terms $\lambda_1$ and $\lambda_2$  are given respectively by
\begin{align}
\lambda_1 = &\frac{1}{3} \sqrt{\frac{ \left( 1- \Omega^2 + \Omega^4 \right)}{\left( 1 - \Omega^2 \right)^{2} } + \frac{3 \times
 \Omega^{4/3}}{ 2^{2/3} \left( -1 + 4\Omega^2-6\Omega^4+4\Omega^6-\Omega^8     \right)^{1/3}}}, \\
\lambda_2=\frac{2}{3} &  \left[ \frac{  \left( 1- \Omega^2+\Omega^4 \right)}{2 \left( 1 - \Omega^2 \right)}
 - \frac{2^{1/3} \times 3 \,  \Omega^{4/3} }{8 \left( -1 + 4 \Omega^2 -6 \Omega^4 + 4 \Omega^6 - \Omega^8   \right) ^{1/3}} \right. \\
& \left. +  \frac{  \left( 2- \Omega^2 \right) \left(1-\Omega^2 - 2 \Omega^4 \right)}{6  \left(1-\Omega^2 \right)^3  \sqrt{\frac{4 \left(
 1- \Omega^2 - \Omega^4 \right)}{9 \left( 1- \Omega^2 \right)^2}  + \frac{2^{4/3} \Omega^{4/3}}{3 \left(-1 + 4 \Omega^2 - 6 \Omega^4 +
 4 \Omega^6 - \Omega^8 \right)^{1/3}} }} \right]^{1/2}.
\end{align}
Using the grand canonical Gibbs free energy
\begin{equation}
G=\frac{\sqrt{\frac{S\left( 1 + S\right) }{1 + S-S \Omega^2}} \left( 1 - S^2 \left( 1- \Omega^2 \right) \right)}{4  \left( 1 + S \right)},
\end{equation}
we get  the Hawking-Page entropy
\begin{equation}
S_{HP}=\frac{1}{\sqrt{1-\Omega^2}}.
\label{Skerrhp}
\end{equation}
A close inspection shows that the calculation  is  quite complicated deserving   appropriate  simplifications.  Replacing Eq.\eqref{Skerrmin}
in Eq.\eqref{TKERR}, we  get   the following formula
\begin{equation}
\begin{aligned}
&  \qquad T_{min} = \frac{\sqrt{3}}{4} \times \lambda_3 \left( \lambda_3 + 2 \sqrt{1- \Omega^2 + \Omega^4}  \right) \times\\
& \frac{1}{\left[\left( 1-2\Omega^2 + \sqrt{1- \Omega^2 + \Omega^4} + \lambda_3  \right) \left( -2+\Omega^2 + \sqrt{1- \Omega^2 + \Omega^4}
 + \lambda_3  \right) \left( 1+\Omega^2 + \sqrt{1- \Omega^2 + \Omega^4} + \lambda_3  \right)\right]^{1/2}},
\end{aligned}
\end{equation}
where the $\lambda_3$ term is
\begin{equation}
\lambda_3=\sqrt{2(1-\Omega^2+\Omega^4)+\frac{(2-\Omega^2)(1-\Omega^2-2\Omega^4)}{\sqrt{1-\Omega^2+\Omega^4}}}.
\end{equation}
The Hawking-Page temperature, associated with  the entropy given in Eq.\eqref{Skerrhp},  reads  as
\begin{equation}
T_{HP}= \sqrt{1-\frac{\Omega^2\left(3+\sqrt{1-\Omega^2} \right)}{4 \left( 1 - \sqrt{1-\Omega^2} \right)}}.
\end{equation}
To obtain the desired  critical ratios, several approximations are needed. For small values of $\Omega$,  we  obtain  the following simplified forms
\bea
 C_S(\Omega)&=& \frac{S_{HP}-S_{min}}{S_{min}}=2 + \frac{27 \Omega^4}{8},  \label{eqrcs} \\
C_T(\Omega)&=& \frac{T_{HP}-T_{min}}{T_{min}}=\frac{2-\sqrt{3}}{\sqrt{3}} + \frac{3\sqrt{3} \Omega^2}{4}.
 \label{eqrct}
\eea
It is easy to recover the previous constants by taking the limit  $\Omega$  goes to zero. At first sight, this  rotation case seems strange.
 A close inspection, however,  shows that this  black hole solution usually exhibits  non-trivial behaviors compared to  the non-rotating one.  Such behaviors
have   been clearly observed   in the investigation of optical aspects  of black holes.  Concretely, it has been shown
  that the shadows of non-rotating black holes processes  a  perfect circle. This
  geometry, however,  has been distorted  by  implementing  the  spin rotation parameters\cite{20,21}. Moreover, circular behaviors can be recovered by taking
   small  values of such parameters. It is  not surprising to have similar ideas for thermodynamic  properties. It is worth noting that the angular velocity dependency in the ratio of Eq\eqref{Skerrmin} by Eq\eqref{Skerrhp} is just analytical. In fact, numerical computations shows that the ratio $S_{min} / S_{HP}$  is equal to the value $1/3$ for small values of the angular velocity. Since analytical expressions are needed,
we have applied several approximations to obtain the ones given in Eq\eqref{eqrcs} and Eq\eqref{eqrct}.

\section{Conclusions and discussions}
Various    numerical  constants  appear  in the formulation of fundamental physical  theories.
    The dimensionless ones
   have been  of a particular interest. The    determination of such scalar constants  is very important. With the help of experimental
evaluations, one  could  predict wether a theory is relevant or not. Besides, the universal constant
 values have been   essential for a precise quantitative description of fundamental theories of the universe. In this paper, we have
   investigated certain    universal constants  arising in   four dimensional
  AdS black holes involving  two primordial critical points. The first critical point corresponds to the Hawking Page phase
   transition described by the entropy $S_{HP}$, while the  second one is associated with the bound under which no black
   hole can survive characterized by the entropy $S_{min}$. These  critical  values produce  $T_{HP}$ and $T_{min}$,
    respectively.  Precisely, we have   shown that the ratios $\frac{S_{HP}}{S_{min}}$ and $ \frac{T_{HP}}{T_{min}}$ can be considered
     as  universal numbers predicted by  a large class of    AdS black hole in four dimensions.  For four dimensional non-rotating AdS black holes,
     we   have found   two  universal constants
       $C_S =\frac{S_{HP}-S_{min}}{S_{min}}=2 $ and  $C_T =\frac{T_{HP}-T_{min}}{T_{min}}= \frac{2-\sqrt{3}}{\sqrt{3}}$.
For the rotating solution,  however, such  universal ratios are slightly affected without losing the expected values.
For  very small  values of the involved rotating parameter, we  have recovered the same
 critical constants.

The present   approach could be
 adaptable for other black holes. In particular, a close examination shows that
   this  analysis might be extended  to non trivial theories of gravity with AdS geometries supported by branes.
A concrete  model  derived  from the compactification of  M-theory  on the sphere $\mathbb{S}^7$ in the presence
 of $N$ coincident $M2$-branes  is   described by the line element $
ds^2=-f(r)dt^2+\frac{1}{f(r)}dr^2  +r^2h_{ij}dx^idx^j+
L^2d\Omega^2_{7} $  where $
f(r)= 1-\frac{m}{r}+ \left( \frac{r}{L} \right)^2$. It is noted that
$h_{ij} \, dx^idx^j$ is the line element of a 2-sphere and  $d\Omega^2_{7} $  is the 7-dimensional sphere with  the raduis $L$
 depending on the brane colors $N^{3/2}$ \cite{14}. Using the reduced entropy $
S=\frac{S_{BH}}{\pi^{3/2}}$ and  $L^9=\frac{\pi^3 N^{3/2}}{2^{3/2}} \ell_p^{9}$,  one  obtains
\begin{equation}
M=   \left[  S^{\frac{1}{2}} \, N^{\frac{7}{12}} + 96 \sqrt{2} \times S^{\frac{3}{2}} N^{-\frac{11}{12}} \right].
\label{M40}
\end{equation}
Using the first law of thermodynamics, one finds the Hawking temperature
\begin{equation}
T=\frac{ \sqrt{3} \pi^{1/6}}{48 \, 2^{7/12} \ell_p} \left[   S^{-\frac{1}{2}} \, N^{\frac{7}{12}} +3
\times 96 \sqrt{2} \, S^{\frac{1}{2}} N^{-\frac{11}{12}} \right].
\label{T40}
\end{equation}
The minimal value of the entropy and the corresponding minimal temperature are
\begin{equation}
\left( S_{min} , T_{min} \right)= \left(  \frac{N^{3/2}}{3^{2}
 \cdot  2^{11/2}},\frac{3^{1/2}\pi^{1/6}}{2^{5/6} \,  N^{1/6} \ell_{p}} \right).
\label{min2}
\end{equation}
Using the grand canonical Gibbs free energy given by
\begin{equation}
G= \frac{ \sqrt{3} \pi^{1/6}}{48 \, 2^{7/12} \ell_p} \left[
  S^{\frac{1}{2}} \, N^{\frac{7}{12}} - 96 \sqrt{2} \times S^{\frac{3}{2}} N^{-\frac{11}{12}} \right],
\label{G40}
\end{equation}
we get   the following critical  point
\begin{equation}
\left( S_{HP}, T_{HP}  \right)=\left(  \frac{ N^{3/2}}{3 \cdot 2^{11/2}}, \frac{2^{1/6} \pi^{1/6}}{ N^{1/6} \ell_{p}}  \right).
\label{HP2}
\end{equation}
Combining Eq.\eqref{min2} and Eq.\eqref{HP2}, we   provide  the  universal critical ratios  of the
 non rotating  case.  According to \cite{14}, these constants  can be also  obtained with DE contributions.\\
 A close inspection  shows that  the two  universal constants  depend on the space-time dimension. To see that,  we consider a concrete
example of  a $d$ dimensional charged AdS black hole given by
  the following  metric
\begin{equation}
f(r)=1-\frac{m}{r^{d-3}}+\frac{q^2}{r^{2 d-6}}+\frac{r^2}{L^2},
\end{equation}
where $q$ is a quantity  related to the charge $
Q=\sqrt{2(d-2)(d-3)} \left( \frac{q}{8 \pi} \right).$
The electrostatic potential conjugated to such a quantity  is $\phi = \frac{\sqrt{d-2}}{2(d-3)} \frac{q}{r_h^{d-3}}$. The calculations show that
 the temperature and the Gibbs free energy for this    charged non-rotating solution
 are given by
\begin{align}
&T =\frac{1}{4 \pi } \left[ \frac{ \omega _{d-2}}{4S} \right]^{\frac{1}{d-2}} \left( (d-3)\left( 1-\frac{2 (d-3)}{d-2}  \phi ^2 \right)+  (d-1) \left[ \frac{ 4S  }{\omega _{d-2}} \right]^{\frac{2}{d-2}}  \right), \\
&G = \frac{1}{ \pi }  \left[ \frac{ S^{d-3} }{ 4^{d-1} \omega _{d-2}} \right]^{\frac{1}{d-2}}  \left( \omega _{d-2}^\frac{2}{d-2} \left( 1-\frac{2 (d-3)}{d-2}  \phi ^2 \right)- \left(  4S  \right)^{\frac{2}{d-2}}  \right).
\end{align}
Using  these expressions,  we obtain
\begin{align}
& S_{min}=\frac{\omega _{d-2} }{4} \left(\frac{d-3}{(d-2)(d-1)}\right)^{\frac{d-2}{2}} \left((d-2)-2(d-3) \phi^2\right)^{\frac{d-2}{2}},   \\
&  T_{min}= \frac{ \sqrt{(d-3) (d-1)} }{2 \pi \sqrt{(d-2) } } \sqrt{(d-2)-2(d-3) \phi^2}, \\
& S_{HP}=\frac{\omega _{d-2} }{4 (d-2)^{\frac{d-2}{2}}} \left((d-2)-2(d-3) \phi^2\right)^{\frac{d-2}{2}}, \\
& T_{HP}= \frac{ \sqrt{(d-2) } }{2 \pi  }  \sqrt{(d-2)-2(d-3) \phi^2}.
\end{align}
Computing  the universal constants for such  a model, we get
\bea
 C_S(d)&=& \frac{ \left( d-3\right)^{\frac{2-d}{2}}-\left( d-1\right)^{\frac{2-d}{2}} }{\left( d-1\right)^{\frac{2-d}{2}}}, \\
C_T(d)&=& \frac{  \left( d-2\right) \sqrt{\left( d-1\right) \left( d-3\right)}-\left( d-1\right) \left( d-3\right) }{\left( d-1\right) \left( d-3\right) } .
\eea
 For $d = 4$, we recover the values given in section 3.

At this point, we remark    that other universal constants related to the angular momentum $J$ or the charge $Q$ of the black hole can be shown to appear in presence of a Hawking-Page transition.
In the extended phase space $(P,V)$,  indeed, there exists an  universal constant by considering  the volume $V$, conjugated variable to  $P$, the Cosmological constant ``pressure''.  In the $d$-dimensional Schwarzschild AdS black hole,  for instance,  calculations provide  the following relations
\begin{align}
 V&= \frac{\left( 4S \right)^{\frac{d-1}{d-2}}}{(d-1)\omega _{d-2}^{\frac{1}{d-2}}}, \label{e2001}\\
S_{min}&=\frac{\omega _{d-2}}{4^{d-1}} \times \left( \frac{(d-3)(d-2)}{\pi \, P} \right)^{\frac{d-2}{2}}, \\
S_{HP}&=\frac{\omega _{d-2}}{4^{d-1}} \times \left( \frac{(d-2)(d-1)}{\pi \, P} \right)^{\frac{d-2}{2}},
\end{align}
where  $\omega _{d-2}$ is the volume of the unit  $(d-2)$-sphere. Combining these relations, we  get  a  new universal constant, related to the cosmological constant
\begin{equation}
\frac{V_{HP}}{V_{min}}=\left( \frac{d-1}{d-3}  \right)^{\frac{d-1}{2}},
\end{equation}
which depends only on  the dimension $d$. Obviously, due to Eq.(\ref{e2001}), this ratio is closely related to the
$C_S$ or $C_T$ quantities.

This work comes up with certain  open questions. One of them is  to  understand such universal
ratio behaviors  for higher dimensional  AdS black holes using shadow analysis. The associated physics  need    deeper reflections and investigations. Another question concerns the variation of external parameters
associated with non-trivial backgrounds on which black hole solutions will be  built.    We leave these questions for future works.
\section*{Acknowledgments}
{\small AB would
like to thank the Departamento de F\'isica, Universidad de Murcia for very kind hospitality
 and scientific supports during the realization of a part of this work. The authors would like to thank  M. Benali, S-E. Ennadifi,  J. J. Fern\'andez-Melgarejo,  H. El Moumni, Y. Hassouni,  K. Masmar,  M. B. Sedra,  and   A. Segui
   for  discussions on related topics.
 The work of ET  has been  supported in part by
the Spanish Ministerio de Universidades and Fundacion
Seneca (CARM Murcia) grants FIS2015-3454, PI2019-
2356B and the Universidad de Murcia project E024-018. This work is partially
supported by the ICTP through AF-13. The authors, being listed in alphabetic order,   are grateful to the anonymous referee for   careful reading of our manuscript, insightful comments, and suggestions, which  improve the present  paper significantly.}


\begin{thebibliography}{10}




  \bibitem{BCH}
  J.~M.~Bardeen, B.~Carter, and S.~W.~Hawking,
  \textit{The four laws of black hole mechanics},
  Commun.\ Math.\ Phys..\ {\bf 31} (1973) 161.

\bibitem{H1}
  S.~W.~Hawking,
  \textit{Black hole explosions},
 Nature\ {\bf 248} (1974) 30.

\bibitem{H2}
  S.~W.~Hawking,
  \textit{Particle creation by black holes},
  Commun.\ Math.\ Phys.\ {\bf 43} (1975) 199.

\bibitem{SV}
  A.~Strominger and C.~Vafa,
 \textit{Microscopic origin of the Bekenstein-Hawking entropy},
  Phys.\ Lett.\ B.\ {\bf 379} (2019) 99.



\bibitem{I1}  S. W. Hawking, W.  Israel, \textit{General Relativity: an Einstein Centenary Survey}, UK Cambridge University Press (2010).



\bibitem{I3} J. B. Hartle, T. Dray, \textit{Gravity: An introduction to Einstein’s general relativity}, Amer. J. Phys. \textbf{71} (2003) 1086.


\bibitem{w2} A. Belhaj, M. Chabab, H. El Moumni, K. Masmar and M. B. Sedra, \textit{Maxwell’s equal-area law for Gauss-Bonnet-Anti-de Sitter black holes}, Eur. Phys. J. C\textbf{75}(2) (2015) 71.

\bibitem{w3} A. Belhaj, M. Chabab, H. El Moumni, K. Masmar, M. B. Sedra, A. Segui, \textit{ On heat properties of AdS black holes in higher dimensions}, JHEP. \textbf{05} (2015) 149.



\bibitem{I9} Y. Liu, D. C. Zou, B. Wang, \textit{Signature of the Van der Waals like small-large charged AdS black hole phase transition in quasinormal modes}, JHEP. \textbf{9} (2014) 179.



\bibitem{w4} A. Belhaj, M. Chabab, H. El Moumni, L. Medari and M. B. Sedra, \textit{The thermodynamical behaviors of Kerr–Newman AdS black holes}, Chin. Phys. Lett. \textbf{30} (2013) 090402.

\bibitem{I11} S. He, L. F. Li, X. X. Zeng, \textit{Holographic Van der Waals-like phase transition in the Gauss-Bonnet gravity}, Nucl. Phys. B\textbf{915} (2017) 243.


\bibitem{I18} D. Kastor, R. Sourya, J. Traschen, \textit{Chemical potential in the first law for holographic entanglement entropy}, JHEP. \textbf{11} (2014) 120.

\bibitem{X0}  R. Maity, P. Roy, T. Sarkar, \textit{Black hole phase transitions and the chemical potential}, Phys. Lett. B\textbf{765} (2017) 386.









 \bibitem{1} S. Hawking, D. N. Page, \textit{Thermodynamics of Black Holes in anti-De Sitter Space}, Commun.Math.Phys. {\bf 87} (1983) 577.

 \bibitem{3}M. Cvetic, S. Gubser, \textit{Thermodynamic stability and phases of general spinning branes}, JHEP {\bf 9907} (1999) 010, {\tt hep-th/9903132}.
  \bibitem{4} A. Chamblin, R. Emparan, C. Johnson,  R. Myers, \textit{Charged AdS black holes and catastrophic holography}, Phys.Rev. D{\bf 60} (1999) 064018,  {\tt hep-th/990217}.
 \bibitem{5}A. Chamblin, R. Emparan, C. Johnson,  R. Myers, \textit{Holography, thermodynamics and fluctuations of charged AdS black holes}, Phys.Rev. D{\bf 60} (1999) 104026, {\tt hep-th/9904197}.




\bibitem{6} S. W. Hawking,  \textit{Black holes in general relativity}, Communications in Mathematical Physics \textbf{25}(2) (1972)  152.


\bibitem{7} J. B. Hartle, T. Dray,  \textit{Gravity: An introduction to Einstein’s general relativity}, Amer. J. Phys. \textbf{71} (2003) 1086.





\bibitem{8} S. W. Hawking, \textit{Black holes and thermodynamics}, Phys. Rev. D\textbf{13}(2) (1976) 191.

\bibitem{9} A. Belhaj, M. Chabab, H. El Moumni and M. B. Sedra, \textit{On thermodynamics of AdS black holes in arbitrary dimensions},
 Chin. Phys. Lett. \textbf{29} (2012) 100401.



\bibitem{10} S. W. Hawking, D. N. Page, \textit{Thermodynamics of black holes in anti-de Sitter space}, Communications in Mathematical
 Physics \textbf{87}(4) (1983) 577.



\bibitem{11} E. Torrente-Lujan, \textit{Smarr mass formulas for BPS multicenter black holes}, Phys. Lett. B\textbf{798} (2019) 135019.



\bibitem{12}  R. Maity, P. Roy, T. Sarkar, \textit{Black hole phase transitions and the chemical potential}, Phys. Lett. B\textbf{765} (2017) 386.

\bibitem{13} A. Belhaj, M. Chabab, H. El Moumni, K. Masmar, M. B. Sedra, \textit{On thermodynamics of AdS black holes in M-theory}, Eur. Phys. J. C\textbf{76}(2)  (2016) 73.

\bibitem{14} A. Belhaj, A. El Balali, W. El Hadri, Y. Hassouni, E. Torrente-Lujan, \textit{Phase
Transitions of Quintessential AdS Black Holes in M-theory/Superstring Inspired Models}, {\tt
arXiv:2004.10647}.

\bibitem{15} S. W. Wei,  Y.X. Liu,  R. B. Mann  \textit{Novel dual relation and constant in Hawking-Page phase transition}, {\tt arxiv.2006.11503}.

\bibitem{16} E. Witten, \textit{Anti-de Sitter space, thermal phase transition, and confinement in gauge theories}, Adv. Theor. Math. Phys. \textbf{2} (1998) 505.

\bibitem{17} D. Kastor, S. Ray, and J. Traschen,  \textit{Enthalpy and the
mechanics of AdS black holes}, Class. Quant. Grav. {\bf 26}
(2009)195011, {\tt arXiv:0904.2765}.

\bibitem{18}A. M. Awad, C. V. Johnson, \textit{Holographic stress tensors for Kerr-AdS black holes},
Phys. Rev. D{\bf 61} (2000) 084025.
\bibitem{19} G. W. Gibbons, M. J. Perry, C. N. Pope,  \textit{The  first law of thermodynamics for
Kerranti-de Sitter black holes}, Class. Quant. Grav. {\bf 22}(2005) 1503.

\bibitem{20} J. M. Bardeen, \textit{Timelike and null geodesics in the Kerr metric}, Les Astres Occlus, (1973).

\bibitem{21} S. Chandrasekhar, \textit{The Mathematical Theory of Black Holes}, Oxford University Press, New York, (1992).

\end{thebibliography}
\end{document}